\documentclass[prl,aps,twocolumn,showpacs,floatfix]{revtex4}
\usepackage{graphicx}
\begin{document}
\title{Geometric Phase and Classical-Quantum Correspondence}
\author{Indubala I. Satija}
\affiliation{Department of Physics, George Mason University, Fairfax, VA 22030}
\author{Radha Balakrishnan}
\affiliation{The Institute of Mathematical Sciences, Chennai  600 113, India}
\date{\today}
\begin{abstract}
{ We study the geometric phase factors underlying 
the classical and the corresponding quantum dynamics of a 
driven nonlinear oscillator 
exhibiting chaotic dynamics. For the classical problem, we
compute the geometric phase factors associated with the 
phase space trajectories using Frenet-Serret formulation. For the 
corresponding quantum problem,
the geometric phase  
associated with the time evolution of the wave function is computed.
Our studies suggest that the classical geometric phase may be related to the
the difference in the quantum geometric phases between two neighboring eigenstates.}
\end{abstract}
\pacs{02.40-k, 05-45.Ac }
\maketitle

Since the discovery of Berry phase\cite{berry}, the study of geometric phases based on the common 
mathematical theme of anholonomy,
has emerged in a variety of fields.\cite{wilczek}
The Berry phase is a path-dependent 
geometric phase associated with the adiabatic time evolution of the  wave function, 
associated with circuits in parameter space.
This concept has been extended\cite{aharanov,MS,sam} to 
non-adiabatic cases and also
to  non-cyclic circuits, since 
the phase acquired by the wave function in  {\it any}
type of time evolution, 
may have a component that is of purely geometric origin.
This phase is a gauge invariant quantity and is equal to the 
difference between the
total phase and the dynamical phase acquired by the wave function.

One of the open questions has been the classical limit of the Berry phase or its generalization
describing the geometric part of the phase of the quantum wave function.
For the special case of integrable Hamiltonian systems, described in terms of action-angle 
variables, the so-called  Hannay angle $\theta_h$\cite{hannay} 
represents the semiclassical limit of the
Berry phase. Berry gave an explicit formula relating the classical angle, called the Hannay angle and the $n^{th}$
quantum eigenstate $\phi_n$ as,
$\theta_h = -\partial_n \phi_n$. 
There have been some attempts to describe the classical limit of Berry phase for chaotic systems
where the effort has been focused on finding a generalization of the {\it 2-form}
within Wigner-Weyl formalism.\cite{RB}

Viewing the Berry phase as an {\it anholonomy effect} underlying dynamical evolution described by
Schr\"{o}dinger equation, 
we seek a classical analog of anholonomy, underlying the corresponding classical evolution
described by the Newton's equations of motion.
Here, we are concerned with the geometric phases 
associated with  periodic, quasiperiodic and
chaotic dynamics of a driven nonlinear oscillator. 
We compute the geometric component of the phase of the wave function 
using a kinematic formulation\cite{MS} of the Berry phase\cite{MS} as given by
Mukunda and Simon. 
 In the corresponding classical
problem, we find the anholonomy underlying nonplanar periodic phase space trajectories and then extend this formulation 
to quasiperiodic and chaotic trajectories.
It should be emphasized that unlike in  Berry phase, our circuits 
are not in parameter space but are in phase space.
By treating a classical phase space trajectory as a space curve, 
we show that  a
geometric phase can be associated with every trajectory, by 
using a Frenet-Serret (FS) formulation\cite{FS}. 
The classical geometric phase is the integrated torsion
of the phase trajectory, and it bears a strong analogy to the geometric 
phase factor associated with the wave function.

To illustrate this correspondence, we begin with the Frenet-Serret equations,
describing the time evolution of the orthonormal FS triad, consisting of the tangent ${\bf T}$, the normal ${\bf N}$, and the
binormal ${\bf B}$ to the space curve ${\bf r}(t)$,
\begin{equation}
\dot{\bf T} = v\kappa {\bf N}\,,\,
\dot{\bf N} = - v\,\kappa \,{\bf T} + v\,\tau \,{\bf B}\,,\,
\dot{\bf B} = - v\,\tau \,{\bf N}\,,
\label{fseqns}
\end{equation}
where $\kappa$ and $\tau$ are respectively the curvature and 
torsion of the curve
and $v = |\dot{\bf r}|$. Here, the overdot denotes derivative with
respect to time.
The above equations can be written as,
$\dot{\bf F}= \xi \times {\bf F}$, where
${\bf F} =
{\bf T},\,{\bf N},$ or ${\bf B},$ and
$\xi =-v\,\kappa {\bf B} + v\tau {\bf T}$.
That is, the FS triad rotates around ${\bf T}$ and ${\bf B}$. One way to
quantify this rotation is to work in a frame in which ${\bf T}$ is
parallel transported, and measure the angle of rotation around the 
tangent 
${\bf T}$. This will be given by the angle 
$\phi_c(t)=\int_{0}^{t} \tau v dt^{\prime}$. Thus $\phi_c(t)$ is 
the geometric phase characterizing
the anholonomy associated with the corresponding phase space orbit.

If we define a complex vector ${\bf M}= ({\bf N}+i{\bf B})/\sqrt {2}$,
 the {\it classical} geometric phase can be written as\\
 $\phi_c(t)= i\int_{0}^{t}
{\bf M}^{*}\cdot \dot {\bf M}~ v dt^{\prime} $.\\
This expression  has exactly the same form as the {\it quantum}
geometric phase found by  Berry \cite{berry}, when  the classical
vector ${\bf M}$ is  replaced by a quantum eigenstate  $\psi_n$.
Additionally, by mapping the closed phase space trajectory to a circuit on a unit sphere
traced by the tip of the tangent vector ( tangent indicatrix), $\phi_c$ can be shown to be the
solid angle subtended by this tangent indicatrix at the center of the
sphere.\cite{radha}
These results are valid also
for a non-periodic trajectory,
since it can  always be closed using a geodesic on the sphere.\cite{MS}

Motivated by this close analogy between the geometric phase in 
a quantum system and the FS geometric phase,
we investigate any possible relationship between the two. 
The underlying key question 
is whether the classical
anholonomy is related to the corresponding quantum one.
Here we calculate the classical and the quantum geometric phases 
for a periodically driven
nonlinear oscillator exhibiting complex dynamics.
The system under investigation is an "impact oscillator"\cite{bs}, 
the oscillator
that rebounds elastically whenever its displacement $x$ drops to zero. 
For $ x > 0$,
the system is described by,
\begin{equation}
H = \frac{1}{2}p^2+\frac{1}{2} \omega_0^2 x^2 -f\cos(\omega t) x
\label{ham}
\end{equation}
The system is piecewise
linear, and the analytic solutions can be obtained for $x > 0$.
The discontinuity at the origin makes it essentially nonlinear.
Without loss of generality, we
choose the units of $t$ and $x$ such that $\omega=1$ and $f=1$.
The phase space trajectory of the dynamical system 
can be viewed as a
space curve generated by the three-dimensional
vector ${\bf r}(t)= ( x, \dot{x}, \ddot{x})$
parameterized by the time $t$. 

As we discuss below, the impact oscillator exhibits very rich and 
complex dynamics,
where periodic, quasiperiodic and chaotic dynamics coexist. 
Our choice of this example was motivated by the fact that in addition to the
simplicity underlying the classical analysis of the oscillator,
the quantum 
wave functions for the driven oscillator are known in terms of the classical solution as,
\cite{Kerner},
\begin{equation}
\psi(x,t)=\chi(x^{\prime}, t)\exp\frac{1}{\hbar}[{\dot{x_c}}(t) x^{\prime}+\int_{0}^{t}L(t^{\prime}) dt^{\prime}]
\end{equation}

Here $x^{\prime} = x-x_c(t)$, with $x_c(t)$ being the solution of the classical equation of motion
and $L$ is the Lagrangian of the driven system. $\chi(x,t)$ is the wave function of the oscillator 
in the absence of driving.
Note that the wave functions of the driven oscillator
are centered on the position of the classical forced oscillator.

If we take $\chi$ to be an eigenstate of the undriven oscillator,
$\chi_n(x,t)= u_n(x) \exp-i(E_n t/\hbar)$,  the eigenfunction 
can be written in terms
of Hermite polynomials as
$u_n(x)=\exp-[\omega_0/(2\hbar) x^2] H_n(\sqrt{\omega_0/\hbar}x)$. 
It should be noted that corresponding to the eigenstate, 
$|\psi(x,t)|^2=|\chi_n(x^{\prime},t)|^2$.
Therefore, the center of the wave packet $x_c(t)$ obeys 
classical equation
of motion, and
the shape of the wave packet ( the density distribution with respect to the center $x_c$)
is unaffected by the driving force.

In view of the fact that we have the analytic solution for the 
classical system for $x > 0$, 
and {\it also} a closed form solution for the quantum wave function, 
we can compute the classical
and the quantum geometric phases with extreme precision. 
These  will be presented below. 

Figures ~1 shows richness and complexity underlying the 
classical dynamics of the oscillator
as we vary the initial energy (initial conditions) of the oscillator.
We see periodic orbits and quasiperiodic tori, in addition to
chaotic trajectories.

All results are for a fixed $\omega_0=1.6$
which corresponds to the oscillator frequency of $3.2$,
in units of $\omega$, the frequency of the driver, putting 
our analysis is close to the adiabatic regime. 

Furthermore, we believe that the probability of inducing a
transition (due to driving) is rather small because the classical
energies of the  particle under consideration here are
far below the threshold for the transition between two quantum
states $n_{1}$ and $n_{2}$ in units of $\hbar \omega_{0}$.

\begin{figure}[htbp]
\includegraphics[width=3.5in]{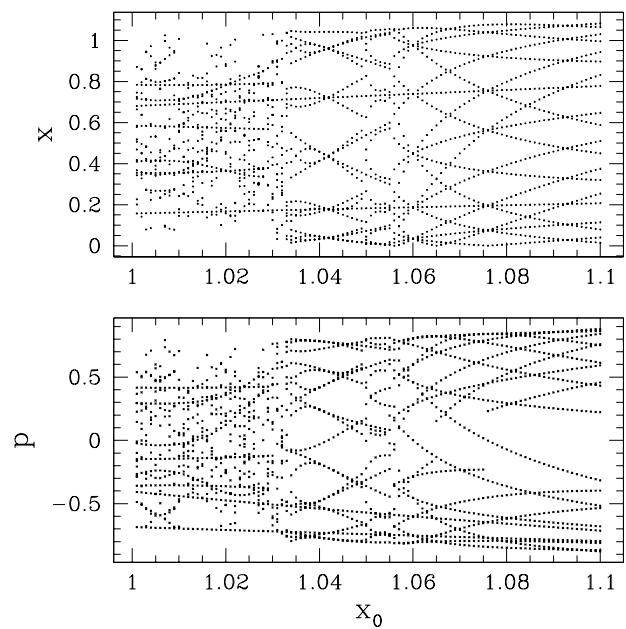}
\leavevmode
\caption{ For  $p_0=0$, the figure shows the 
possible $x$ and $p$ 
values (once every period of the driver), as a function 
of $x_0$, the initial position 
of the oscillator.
For $x_0 \approx 1.0416$, there is a period-$11$ orbit. 
For $x_0 > 1.046$,
we get invariant quasiperiodic tori or cantori as seen in 
the lower figure. For $x_0 < 1.035$, almost all initial
conditions result in chaotic dynamics.}
\label{fig1}
\end{figure}

Using the classical solution, the FS geometric phase $\phi_c$ 
can be computed as the integral of the torsion, given by
$\tau= {\bf r}_{t} \cdot({\bf r}_{tt}
\times {\bf r}_{ttt}) /|{\bf r}_{t}\times {\bf r}_{tt}|^2$
For the corresponding quantum problem, 
we can obtain the geometric phase,
using kinematic formulation by Mukunda and Simon\cite{MS}.
For a given wave function $\psi(x,t)$, the quantum geometric phase $\phi_q$is a gauge invariant quantity is given by
$\phi_q(t)=arg\int[\psi^*(x,0) \psi(x,t)]dx-Im\int_{0}^{t}
\psi^*(x,t^{\prime})\partial_t\psi(x,t^{\prime})dt^{\prime}$.
It is clear that the first term describes total phase while the second term describes the dynamical phase
accumulated by the wave function in time $t$.
Note that this formulation does not require any circuits to define geometric phase.

For the driven impact oscillator, if we substitute the solution for the wave function, given by equation (3).
, the geometric phase can be written in terms of the expectation values of $<x>$ and the classical solution $x_c$ ,
\begin{equation}
-\hbar \phi_{q,n}=\int_{0}^{t} [L(t^{\prime})-<x> \ddot{x_c}(t^{\prime})]dt^{\prime}+G(t)
\end{equation}
where $G(t)$ is given by,
$G(t)=(\dot{x_c}(t)x_c(t)-\dot{x_c}(0)x_c(0))+arg[\int[(u_n(x^{\prime}(0)u_n(x^{\prime}(t)
\exp\frac{ix}{\hbar}(\dot{x_c}(t)-\dot{x_c}(0))]dx$.
For periodic evolution, $G(t)=0$. Also, $G(t)=0$ if we consider $t$ values where the classical
particle is at the turning points. In view of this, we will confine our calculations of geometric phases
to only such values of $t$.

Since the classical impact oscillator is confined to $x \ge 0$, the eigenstates of the corresponding quantum
system are restricted  to  {\it odd} $n$ values only. Substituting the explicit expression for the Hermite polynomials,
$<x>$ can be expressed in terms
of parabolic cylindrical functions which are functions of $x_c$.
\begin{equation}
<x>_{2n-1} = \frac{\sum_{l=1}^{4n} A_{4n-l}(y_c) D_{-l}}{\sum_{l=1}^{4n-1} B_{4n-l}(y_c) D_{-l}}
\end{equation}
where $D_{-m}(y_c)$ is a parabolic cylinder functions of order $m$.
Here $y_c=\alpha x_c$ where $\alpha=\sqrt{2\omega_0/\hbar}$. 
$A_l(y_c)$ and $B_l(y_c)$ are polynomials of $y_c$ of degree $l$ that are uniquely determined by $n$.
For example, for $n=1$, we have
\begin{equation}
<x>_{1}=\frac{1}{\alpha}\frac{6D_{-4}(y_c)-4y_cD_{-3}(y_c)+y_c^2D_{-2}(y_c)}{2D_{-3}(y_c)-2y_cD_{-2}(y_c)+y_c^2D_{-1}(y_c)}
\end{equation}

For higher values of $n$, the expressions are very complicated. 
Our analysis has been confined to the ground state $n=1$, and the
first excited state $n=3$. For all values of $n$,
$\hbar\phi_{q,n}$ reduces to the classical action as $\hbar \to 0$. 
We factor out this part
of the phase factor $\phi_0$ and 
write $\phi_{q,n}=\phi_0+\phi_n$. As we show below, it is the
difference in the phases between the two neighboring eigenstates that exhibit quantum fingerprints of classical dynamics.

Figures ~2 and ~3 show geometric phases for a fixed time
interval $T$ , equal to the driving period, for a periodic and a chaotic trajectory.
These results as well as other similar analysis suggest that $\phi_c$ may be related to $\phi_{n-1}-\phi_n$.

For dynamical evolution involving arbitrary time $t$, our detailed analysis shows that the integrated quantum phase factors
can be expressed as a sum of linear and oscillatory functions of $t$ and hence can be written as
, $\phi_n(t)=v_nt+f_n(t)$. Here
$v_n$ are constants, independent of $t$ and $f_n(t)$ are oscillatory functions which are found to
retain the fingerprints of the corresponding classical dynamics for all times.

In contrast to quantum phases, the classical $\phi_c(t)$ is found to be an oscillatory function
for periodic , quasiperiodic as well as for the chaotic trajectories.
This should be contrasted with the the driven and damped oscillator 
phases\cite{QP1} where the classical phase
averaged over the period of the driver was finite.
\begin{figure}[htbp]
\includegraphics[width=3.5in]{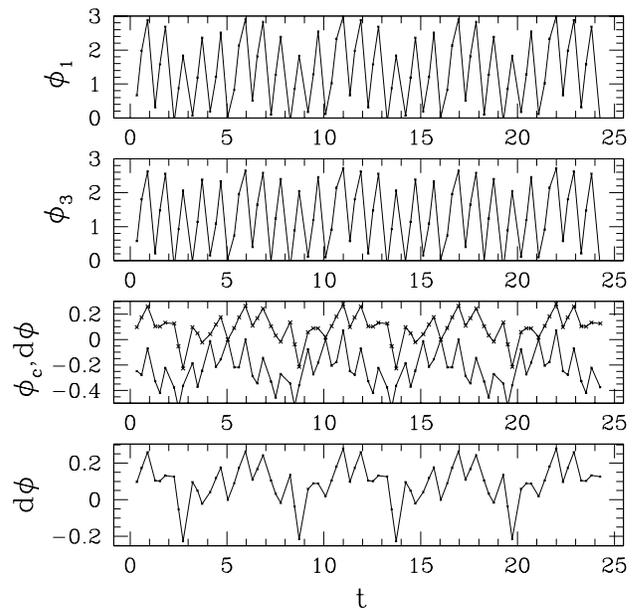}
\leavevmode
\caption{ For a period-$11$ trajectory, the four figures 
(from top to bottom) show  the quantum geometric phases for 
the ground state( $n=1$), the first excited state ( $n=3$), 
classical geometric phase $\phi_c$ and $d\phi=\phi_1-\phi_3$,
respectively.}
\label{fig2}
\end{figure}

\begin{figure}[htbp]
\includegraphics[width=3.5in]{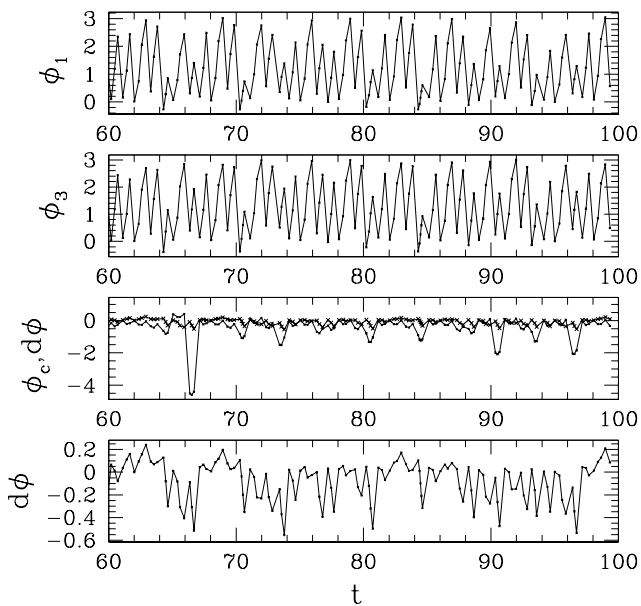}
\leavevmode
\caption{For a chaotic trajectory, same plots as in Fig. ~2 
In the third figure, we superimpose $d\phi$ with $\phi_c$.} 
\label{fig3}
\end{figure}

The results for arbirary time evolution are shown in
figures ~4 and ~5. Here we factor out the linearly increasing 
parts of the phases 
and show the fluctuating parts along with the
classical phase. For the parameter values corresponding to 
the figures 2 to 5,
$v_1/(2\pi)=.4218$ and $v_3/(2\pi)=.3956$ for
the periodic orbit and $v_1/(2\pi)=.3723$ and $v_3/(2\pi)=.4021$ 
for the chaotic orbits.
We notice that $\phi_c$ is of the same order of magnitude 
as $f_1-f_3$ with $\phi_c$
exhibiting intermittent fluctuations. 
In view of the fact that $v_1 \approx v_3$, it is 
conceivable that $v_n$ approaches a constant, independent
of $n$ as $n \rightarrow \infty$
and therefore $\phi_c$ may be related to the $d\phi$ for 
arbitrary time $t$.
This is analogous to Berry's relation $\theta_n = -\partial_n 
\psi_n$ which was true for integrable systems in semiclassical limit.

\begin{figure}[htbp]
\includegraphics[width=3.5in]{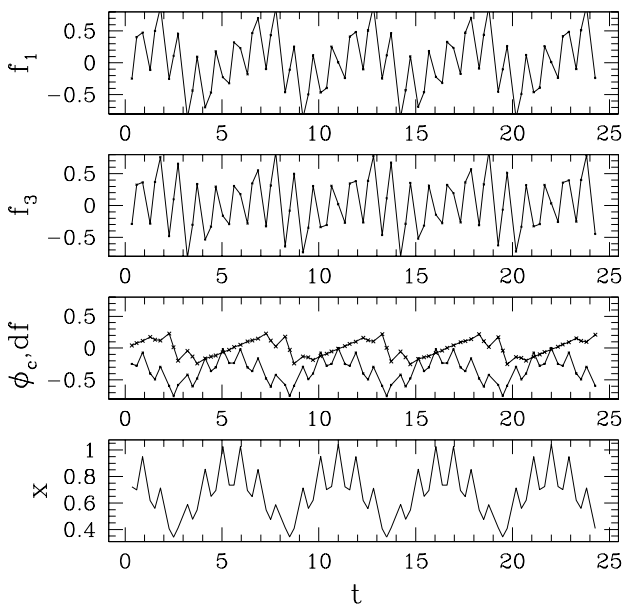}
\leavevmode
\caption{For a period-11 trajectory, the top two figures 
show the oscillatory 
components $f_1$ and $f_3$ of the net accumulated phases
for the ground state and the excited state. 
The third figure shows the classical phase
and the differnce $df=f_1-f_3$. The lowest figure 
shows the position of the 
oscillator. }
\label{fig4}
\end{figure}

\begin{figure}[htbp]
\includegraphics[width=3.5in]{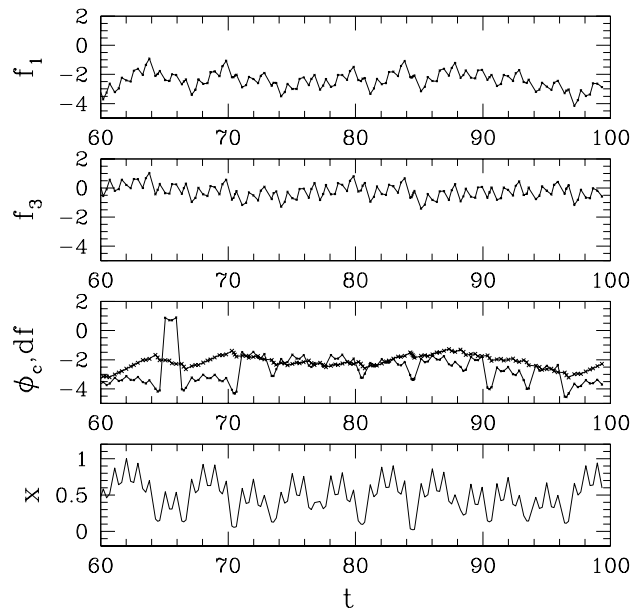}
\leavevmode
\caption{For a chaotic trajectory, the same plots as in 
Fig. ~4. }
\label{fig5}
\end{figure}

In summary, our results as depicted above describe preliminary 
studies of classical and quantum phases
underlying a driven nonlinear oscillator. An interesting 
result is the possible relationship
between the classical phases and the difference in the 
quantum phases between the two
neighboring states, reminiscent of the Berry relation 
$\theta_h=-\partial_n \phi_n$.
As a consequence, the smalless of the classical phase 
will have its origin in
the relative stability of the quantum phase with respect to 
the quantum number $n$.
It is rather intriguing that the fluctuations in the quantum 
geometric phases retain the finger prints of the corresponding 
classical dynamics.
Further detailed studies involving higher excited states 
are needed to confirm these speculative views.

In physical applications such as atom optics, Hamiltonians of the form,
$H(x,t)=H_0+\lambda x \sin(\omega t)$ are relevant where $H_0$ 
is the time-independent
Hamiltonian and
the time-dependent term describes
the interaction with a single mode radiation field in 
dipole approximation.
For nonlinear $H_0$, nontrivial dynamics may lead to 
many surprises. The impact oscillators
exhibit dynamics with many features that are typical of a 
nonlinear oscillator.
This suggests that the driven impact oscillator may provide 
an interesting theoretical model
to explore various issues relevant to quantum chaos.

The research of IIS is supported by National Science
Foundation Grant No. DMR~0072813.

\end{document}